%% file: main.tex
\documentclass{llncs}
\usepackage{makeidx}  
\usepackage{color,graphicx}
\usepackage{graphicx}

\begin{document}

\frontmatter          
\pagestyle{headings}  
\addtocmark{ }        

\mainmatter           

\title{ Are we there yet?  When to stop a Markov chain while
generating random graphs\thanks{This work was funded by the Applied
Mathematics Program at the U.S.\@ Department of Energy and performed
at Sandia National Laboratories, a multiprogram laboratory operated by
Sandia Corporation, a wholly owned subsidiary of Lockheed Martin
Corporation, for the United States Department of Energy's National
Nuclear Security Administration under contract DE-AC04-94AL85000.}}

\titlerunning{Independent graph generation}

\author{Jaideep Ray \and Ali Pinar \and C. Seshadhri}
\authorrunning{J. Ray et al.} 

\tocauthor{J. Ray, A. Pinar and C. Seshadhri}

\institute{Sandia National Laboratories, Livermore, CA 94550 \\
\email{\{jairay,apinar,scomand\}@sandia.gov}}

\maketitle              
\let\thefootnote\relax\footnotetext{\emph{Submitted to the 9th Workshop on Algorithms and Models 
for the Web Graph, WAW-2012, June 22-23, 2012, Dalhousie University, Halifax, Nova Scotia, Canada. SAND2012-1169C}}

\begin{abstract}
Markov chains are a convenient means of generating realizations of
networks, since they require little more than a procedure for rewiring
edges. If a rewiring procedure exists for generating new graphs with
specified statistical properties, then a Markov chain sampler can
generate an ensemble of graphs with prescribed
characteristics. However, successive graphs in a Markov chain cannot
be used when one desires independent draws from the distribution of
graphs; the realizations are correlated. Consequently, one runs a
Markov chain for $N$ iterations before accepting the realization as an
independent sample. In this work, we devise two methods for
calculating $N$. They are both based on the binary ``time-series''
denoting the occurrence/non-occurrence of edge $(u, v)$ between
vertices $u$ and $v$ in the Markov chain of graphs generated by the
sampler. They differ in their underlying assumptions. We test them on
the generation of graphs with a prescribed joint degree
distribution. We find the $N \propto |E|$, where $|E|$ is the number
of edges in the graph. The two methods are compared by sampling on
real, sparse graphs with ${\rm 10^3} - {\rm 10^4}$ vertices.
\end{abstract}

\input{defn}

\input{intro}

\input{jdd}


\input{models}

\input{tests}

\input{concl}

\bibliographystyle{splncs} 
\bibliography{jdd,main}
\end{document}

%% file: defn.tex

\newcommand{\M}[1]{\mathbf{#1}} 

\newcommand{\iat} {$\tau_{int} $}
\newcommand{\bv}  {{\mathbf v}}
\newcommand{\bp}  {{\mathbf p}}
\newcommand{\bu}  {{\mathbf u}}
\newcommand{\MC}  {${\mathcal M}$}
\newcommand{\be}  {{\mathbf e}}
\newcommand{\eps} {\epsilon}

\newcommand{\Comment}[2] {\emph{\color{#1} #2}}

%% file: intro.tex
\section{Introduction}
\label{sec:intro}

Markov chain Monte Carlo (MCMC) methods are a common means of
generating realizations of graphs which share similar characteristics
since they require nothing more than a procedure that can generate a
new graph by ``rewiring'' the edges of an existing graph. Much of
their use to date has been in generating graphs with a prescribed
degree
distribution~\cite{ktv,JerrumS90,JerrumSV04,GkantsidisMMZ03}. Other
efforts have used MCMC to generate graphs with a prescribed joint
degree distribution~\cite{StPi12}.  MCMC methods require a graph to
start the chain; thereafter, the ``rewiring'' procedure generates new,
realizations which preserve certain graph characteristics. The
specific characteristic(s) that are preserved depend entirely on the
``rewiring'' procedure.

MCMC methods for generating graphs has two drawbacks. \emph{Initialization bias} arises from the fact that the
starting graph may not even lie in the population of graphs that we
seek to sample, or may be an outlier in that population. The second
issue, \emph{autocorrelation in equilibrium}, arises from the fact
that successive samples drawn by the MCMC sampler are correlated and
an empirical distribution constructed from them would result in the
statistical error (variance) to be 2\iat larger than the distribution
constructed using independent samples. Here \iat, the integrated
autocorrelation time, is a measure of how slowly correlation in
graphical metrics, calculated from an MCMC series, decays; see Sections
2 and 3 in Sokal's lecture notes~\cite{sokal}.

Sokal's method~\cite{sokal} for deciding the ``sufficiency'' of
samples obtained from MCMC revolve around autocorrelation. The method
is general, and was adapted for use with graphs
in~\cite{StPi12}. Consider an edge $(u, v)$ between labeled vertices
$u$ and $v$ in the ensemble of graphs generated by the MCMC
chain. Denoting its occurrence/non-occurrence in the chain of graphs
by 1/0 gives us a binary time-series $\{Z_t\}, t = 1 \ldots T$, with
an empirical mean $\mu$. The auto-correlation, with lag $l$, is given
by $C(l) = (\{Z_t\} - \mu) ( \{Z_{t+l}\} - \mu ), t = 1 \ldots T-l$
and the normalized version of it, $\rho(l) = C(l) / C(0)$ can be used
to gauge whether the autocorrelation in the time-series is observed to
be decreasing. In~\cite{StPi12}, the authors used this metric, applied
to all edges in the graphs that were sampled, to ensure that their
MCMC chain was mixed.  One can also set, loosely speaking (for
details, consult~\cite{sokal}), a minimum threshold $\rho_{min}$,
identify the corresponding lag $l_{min}$, and retain every
$l_{min}^{th}$ entry in the MCMC chain to serve as independent
samples.  However, this method has two practical drawbacks. First the
autocorrelation analysis has to be performed for all the edges
(potentially, $|V|^2$ in number) that might appear in the MCMC chain,
which quickly becomes prohibitively expensive for large
graphs. Secondly, it requires a user input, $\rho_{min}$, which may
have an arbitrary effect on graphical properties of the
ensemble. These shortcomings motivate our work.

In this paper, we propose two different methods for generating
\emph{independent} graphs using an MCMC method. The first, which we
call Method A or ``multiple short runs'', determines the number of
iterations $N$ an MCMC method has to be run to ``forget'' the initial
graph and minimize the initialization bias.  The second approach,
Method B or ``one long run'', requires $K$ MCMC iterations. This long
run is thinned by a factor $k$ (i.e., every $k^{th}$ MCMC iteration is
preserved) to generate $K/k$ independent samples. Both methods are
intended to be approximate, but simple to evaluate, so that they can
be employed in practice to gauge the ``sufficiency'' of MCMC
iterations. The two methods for extracting independent graphs are
tested on an MCMC chain with the setup described in~\cite{StPi12}. We
explore the practical impact of approximations in our methods. We
restrict ourselves to undirected graphs with labeled nodes.

In the next section (Sec.~\ref{sec:jdd}) we describe the procedure
used to ``rewire'' a graph to create a new graph realization with the
same joint degree distribution. In Sec.~\ref{sec:methods} we describe
the two methods for generating independent samples. In
Sec.~\ref{sec:tests}, we test the methods on real sparse graphs. We
conclude in Sec.~\ref{sec:concl}.

%% file: jdd.tex
\section{A  Markov chain algorithm for sampling graphs with 
a given joint  degree distribution}
\label{sec:jdd}

Consider an undirected graph $G=(V,E)$, where $|V| = n$ and $|E| = m$.
The degree distribution of the graph is given by the vector $\vec{f}$,
where $f(d)$ is the number of vertices of degree $d$.  The \emph{joint
degree distribution} list the number of edges incident between
vertices of specified degrees. Formally, the $n \times n$ matrix
$\M{J}$ denotes the joint degree distribution, where the entry
$J(i,j)$ is the number of edges between vertices of degree $i$ and
degree $j$. Stanton and Pinar~\cite{StPi12} studied the problem of
generating and random sampling a graph with a given joint degree
distribution. They proposed a greedy algorithm to construct an
instance of a graph with a specified (feasible) degree distribution,
as well as a Markov chain algorithm to generate random samples of
graphs with the same degree distribution.
 
\begin{figure}
\centerline{\includegraphics[width=\textwidth] {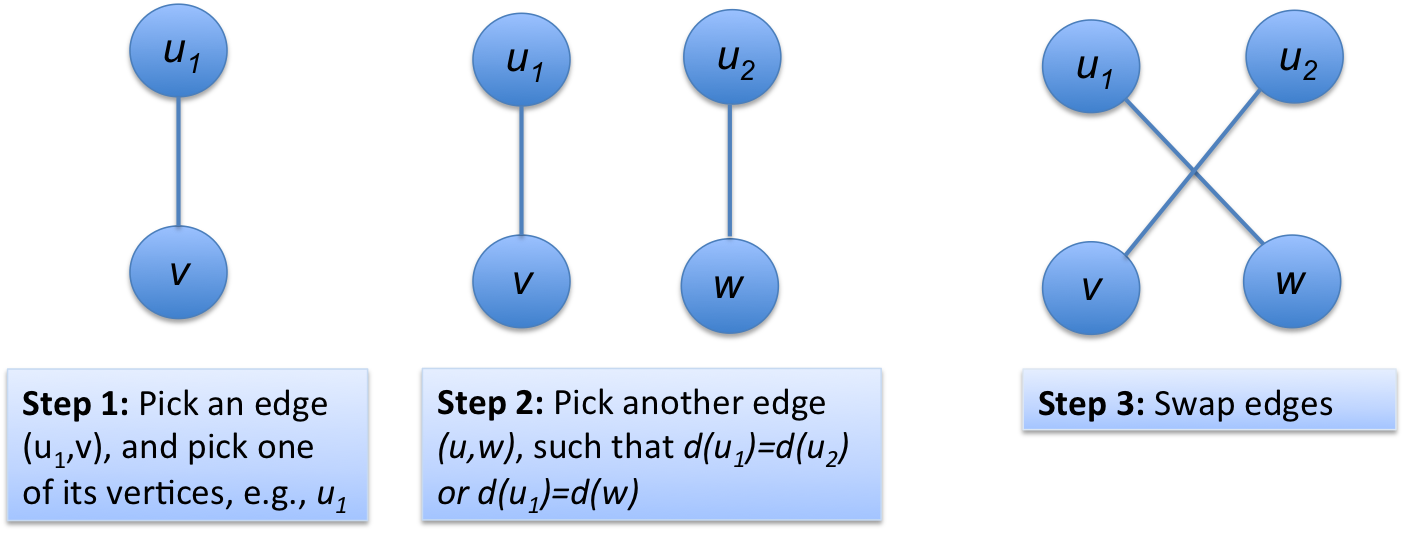}}
\caption{\label{fig:jdd} The swapping operation for the Markov chain
algorithm.}
\end{figure}

For the purposes of this paper, we will only focus on the Markov chain
algorithm.  The rewiring operation that moves us between the nodes of
the Markov chain is depicted in Fig.~\ref{fig:jdd}. At the first step,
one picks an edge $(u_1, v)$ at random and thereafter, one of the end
vertices, e.g., $u_1$. We wish to break $(u_1, v)$ and connect $u_1$
and $v$ to others without violating the prescribed joint degree
distribution. We, therefore, search for another edge $(u_2, w)$ where
$d_{u_2} = d_{u_1}$ or $ d_w = d_{u_1}$, where $d_p$ denotes the
degree of node $p$. WLOG, let $d_{u_2} = d_{u_1}$. Swapping the edges
i.e. creating edges $(u_1, w)$ and $(u_2, v)$ while destroying $(u_1,
v)$ and $(u_2, w)$ leaves the joint degree distribution unchanged
while changing the connectivity pattern of the graph. If the resulting
graph is simple, the graph is retained by the MCMC chain.

The procedure results in inter-edge correlation. A particular edge is
chosen for swapping, on average, once every $|E|$ Markov chain
iterations. In~\cite{StPi12}, this procedure was used to generate
graphs (of moderate sizes, with $|V| \leq 23,000$), using an MCMC
sampler. Autocorrelation analysis showed that the Markov chain mixes
and the autocorrelation decays with edge-dependent rates. Empirically
it was observed that an edge $(u, v)$ de-correlated slowly if $J(d_u,
d_v) / (f{d_u} f{d_v}) \approx 0.5$.  Empirically, it was observed
that the autocorrelations of the edges decreased very sharply.

%% file: models.tex
\section{Methods for calculating independence}
\label{sec:methods}

 The algorithm in Sec.~\ref{sec:jdd} has two variants - one where the
resulting graph may not be simple, and another where the graph was
always simple (A graph is simple if it does not have self-loops or
parallel edges).  For the first variant, earlier work~\cite{ktv}
implies that this chain has a polynomial mixing time.  The second
variant is not even known to have any bounds on the mixing time.  As
discussed in Sec.~\ref{sec:intro}, one would like to provide a length
to run the MCMC that, although not guaranteeing complete mixing, at
least gives some confidence that the sampled graph is fairly
``random." To that end, we will we will approximate the behavior of a
single edge by a Markov chain. We stress that we do not give a proof,
but only a mathematical argument justifying this. Below we present two
methods for generating independent graph samples.

\subsection{Method A - ``multiple short runs''}
\label{sec:sesh}

We refer to this method as Method A or the ``multiple short runs''
method. We generate $M$ graph samples by running $M$ independent
Markov chain for $N$ iterations before accepting the resulting
graph. All the chains are run from the same initial graph; however,
the state in the random number generator in each of the $M$ MCMC
chains are distinct.

Consider a fixed pair of labeled vertices $\{u, v\}$. We will
approximate the occurrence of an edge between $(u,v)$ as a two state
Markov chain. Note that this is an approximation, since these
transitions depend on the remaining graph if $\M{J}$ is to be
preserved. Nonetheless, these dependencies appear to be weak. The
first coordinate of the matrix is state $0$ (no edge) and the second
coordinate is $1$, indicating the existence of an edge. The transition
matrix $\M{T}$ for this chain is 
\begin{equation}
  \M{T}_{i,j} = \left( \begin{array}{cc}
                      1 - \alpha_{i,j}   &   \alpha_{i,j} \\
                      \beta_{i,j}        & 1 - \beta_{i,j}
                       \end{array} \right),
\label{eqn:transition}
\end{equation}
where $i = d_u$ and $j = d_v$ are the degrees of vertices $u$ and
$v$. $\alpha_{i,j}$ and $\beta_{i,j}$ are positive fractions and
$(\alpha_{i,j} + \beta_{i,j}) \leq 1$. The eigenvalues of the
transition matrix are 1 and $ 1 - (\alpha_{i,j}+\beta_{i,j})$. Below,
we construct a model for $\alpha_{i,j}$ and $\beta_{i,j}$.

Suppose the state is currently $0$. The state will become $1$ if the
edge $(u,v)$ is swapped in. Let the two edges chosen by the algorithm
be $e$ and $e'$, in that order. The edge $(u,v)$ is swapped in if $e$
contains $u$ and $e'$ contains $v$ (or vice versa). Furthermore, the
endpoint $u$ must be chosen, and the other end of $e'$ must have
degree $d_u$.  The probability that $e$ contains $u$ and $u$ is chosen
as an endpoint is exactly $d_u/2m$.  The probability we choose the
edge $e'$ that is incident to $v$ depends on the number of neighbors
of $v$ whose degree is $d_u$. Clearly, this depends on the graph
structure (leading in a non-Markov probability of this
transition). We heuristically guess this number based on the joint
degree distribution. The number of edges from degree $d_u$ to degree
$d_v$ vertices is $\M{J}(d_u,d_v)$. Of these, the average number of
edges incident to a fixed vertex of degree $d_v$ is
$\M{J}(d_u,d_v)/f_v$. We shall approximate the number of edges
incident to $v$ with the other endpoint of degree $d_u$ by this
quantity. The total probability is
\[ \frac{d_u}{2m} \times \frac{\M{J}(d_u,d_v)}{mf(d_v)} = \frac{d_u \M{J}(d_u,d_v)}{2m^2f(d_v)} \]
The edge $(u,v)$ is also swapped in when the reverse happens (so we choose $v$ as an endpoint,
and an edge incident to $u$ with the other endpoint of degree $d_v$). The total
transition probability from $0$ to $1$ is approximated by
\begin{equation}
  \alpha_{i,j} =   \frac{d_u \M{J}(d_u,d_v)}{2m^2f(d_v)} 
                 + \frac{d_v \M{J}(d_u,d_v)}{2m^2f(d_u)} 
               =   \frac{\M{J}(d_u,d_v)}{2m^2}
                   \left(\frac{d_u}{ f(d_v)} + \frac{d_v}{f(d_u)}\right)
\label{eqn:alpha}
\end{equation}

We now address the transition from $1$ to $0$. Suppose $(u,v)$ is
currently an edge.  If the first edge $e$ is chosen to be $(u,v)$,
then $(u,v)$ will definitely be swapped out. The probability of this
is $1/m$. If the random endpoint chosen has degree $d_u$ (and is not
$u$), then we might choose $e'$ to be $(u,v)$. The total probability
of this is
\[  \frac{(f(d_u) - 1)d_u}{2m} \times \frac{1}{m} = \frac{(f(d_u) - 1)d_u}{2m^2} \]
The roles of $u$ and $v$ can also be reversed, so the total transition
probability from $1$ to $0$ is
\[ \frac{f(d_u)d_u + f(d_v)d_v - d_u - d_v}{2m^2} \]
and so
\begin{equation}
     \beta_{i,j} = \frac{1}{m} + \frac{f(d_u)d_u + f(d_v)d_v - d_u -  d_v}{2m^2}
\label{eqn:beta}
\end{equation}

We proceed to determining the number of iterations $N$ to run the
Markov chain. We start the Markov chain \MC \hspace{2mm} with an
initial distribution $\bv$ (which is either $(0, 1)$ or $(1,
0)$). \MC, which is represented by $\M{T}_{i,j}$
(Eq.~\ref{eqn:transition}), is run for $N = ln(1/\epsilon) / (\alpha +
\beta)$ iterations, $\epsilon > 0$. We have dropped the subscripts $i$
and $j$, since it is implied that this model is being derived for an
edge $(u,v)$ with vertices of degrees $i$ and $j$. After $N$ steps, we
realize a 2-state distribution $\bp = (p_0, p_1)$, which is different
from the stationary distribution be $\bu = (u_0, u_1)$.

Denote the unit 2-norm eigenvectors of $\M{T}$, corresponding to the
eigenvalues 1 and $1 - (\alpha+\beta)$, as $\be_1$ and $\be_2$. The
initial state can be expressed as $\bv = c_1 \be_1 + c_2 \be_2$. After
$N$ applications of the transition matrix we get
\[ \bp = \M{T}^N \bv = c_1 \M{T}^N \be_1 + c_2 \M{T}^N \be_2 
       = c_1 \be_1 + c_2 \left(1 - (\alpha+\beta) \right)^N \be_2. \]
Since $(1 - \{\alpha+\beta\}) < 1$, the second term decays with $N$
and $c_1 \be_1$ is the stationary distribution. We can bound the
decaying term as
\[
 \|(1-(\alpha+\beta))^N c_2 \be_2\|_2 = (1-(\alpha+\beta))^{\ln(1/\eps)/(\alpha + \beta)} c_2 \|\be_2\|_2
\leq \exp(-\ln(1/\eps)) = \eps 
\]
Hence, $\|\bp-\bu\|_2 \leq \eps$, and so each $|p_i - u_i|$ is at most
$\eps$. Further, from Eq.~\ref{eqn:alpha} and \ref{eqn:beta}, we see
that $\alpha + \beta \geq 1/m$ (to leading order) and consequently
\begin{equation}
N = \frac{\ln(1/\eps)}{\alpha + \beta} \leq m \ln(1/\eps) 
= |E| \ln\left(\frac{1}{\epsilon}\right).
\label{eqn:modela}
\end{equation}

\subsection{Method B - ``one long run''}
\label{sec:rl}

We propose a second method, which we refer to as Method B or ``one
long run'', for generating independent graphs. The procedure involves
running a Markov chain for a large number of steps $K$ and thinning
it by a factor $k$ i.e., preserving every $k^{th}$ instance of the
chain. Comparing with the development in Sec.~\ref{sec:sesh}, we
expect $k \sim N$.

Similar to Method A (Sec.~\ref{sec:sesh}), this method too begins with
the binary time-series of edge occurrence $\{Z_t\}$. As observed in
~\cite{StPi12}, the autocorrelation in $\{Z_t\}$ decays for all
edges. Consequently it is possible to successively thin the chain
$\{Z_t\}$ (i.e., retain every $k^{th}$ element to obtain $\{Z^k_t\}$,
the $k-$thinned chain) and compare the likelihoods that the chains
were generated by (1) independent sampling or (2) by a first-order
Markov process. When sufficiently thinned, the independent sampling
model is expected to fit the data better. Using this as the stopping
criterion removes an ambiguity (user-specified tolerances).  We will
employ a method based on comparison of log-likelihoods of model
fit. We derive these expressions below. While this technique has been
applied in other domains~\cite{96ra1a,92rl2a}, but this paper is the
first application of this technique to graphs.

Consider the chain $\{Z^k_t\}$. We count the number, $x_{ij}$, of the
$(i, j), i, j \in (0, 1)$ transitions in it. $x_{ij}$ are used to
populate $X$, a $2 \times 2$ contingency table. Dividing each entry by
the length of thinned chain $K/k -1$ provides us with the empirical
probabilities $p_{ij}$ of observing an $(i,j)$ transition in
$\{Z^k_t\}$. Let $\widehat{p_{ij}}$ and $\widehat{x_{ij}} = (K/k -1)
\widehat{p_{ij}}$ be the predictions of the probabilities and expected
values of the table entries provided by a model. In such a case, the
goodness-of-fit of the model is provided by a likelihood ratio
statistic (called the $G^2$-statistic; Chapter 4.2 in~\cite{07bfh3a})
and a Bayesian Information Criterion (BIC) score
\begin{equation}
  G^2 = -2 \sum_{i=0}^{i=1} \sum_{i=0}^{i=1} x_{ij} \log\left( \frac{\widehat{x_{ij}}}{x_{ij}}\right), 
  \mbox{\hspace{1cm}} BIC = G^2 + n \log\left( \frac{K}{k} - 1,
  \right)
\label{eqn:ll}
\end{equation}
where $n$ is the number of parameters in the model used to fit the
table data. Typically log-linear models are used for the purpose
(Chapter 2.2.3 in~\cite{07bfh3a}); the log-linear models for table
entries generated by independent sampling and a first-order Markov
process are
\begin{equation}
\log(p_{ij}^{(I)}) = u^{(I)} + u^{(I)}_{1,(i)} + u^{(I)}_{2, (j)} 
\mbox{\hspace{1mm} and \hspace{1mm}}
\log(p_{ij}^{(M)}) = u^{(M)} + u^{(M)}_{1,(i)} + u^{(M)}_{2, (j)} + u^{(M)}_{12,(ij)},
\label{eqn:llm}
\end{equation}
where superscripts $I, M$ indicate an independent and Markov process
respectively. The maximum likelihood estimates (MLE) of the model
parameters ($u^{(W)}_{b,(c)}$) are available in closed form (Chapter
3.1.1 in~\cite{07bfh3a}). They lead to the model predictions below
\begin{equation}
\widehat{x_{ij}^{I}}  =  \frac{(x_{i+}) (x_{+j})}{x_{++}} 
\mbox{\hspace{2mm} and \hspace{2mm}}
\widehat{x_{ij}^{M}}  =  x_{ij},
\label{eqn:llmans}
\end{equation}
where $x_{i+}$ and $x_{+j}$ are the sums of the table entries in row
$i$ and column $j$ respectively. $x_{++}$ is the sum of all entries
(i.e., $K/k - 1$, the number of transitions observed in $\{Z^k_t\}$,
or the total number of data points). We compare the fits of the two
models thus:
\begin{equation}
 \Delta BIC =  BIC^{(I)} - BIC^{(M)} 
            =  -2 \sum_{i=0}^{i=1} \sum_{i=0}^{i=1} x_{ij}
              \log\left( \frac{\widehat{x_{ij}^{(I)}}}{x_{ij}}\right)
              - \log\left(\frac{K}{k} - 1\right).
\label{eqn:bic}
\end{equation}
Above, we have substituted $\widehat{x_{ij}^{(M)}} = x_{ij}$ and the
fact that the log-linear model for a Markov process has one more
parameter than the independent sampler model. Large BIC values
indicate a bad fit. A negative $\Delta BIC$ indicates that an
independent model fits better than a Markov model. 

The procedure for identifying a suitable thinning factor $k$ then
reduces to progressively thinning $\{Z^k_t\}$ till $\Delta BIC$ in
Eq.~\ref{eqn:bic} becomes negative. We search for $k$ in powers of
2. The value of $k$ so obtained varies between edges and
conservatively, we take the largest $k, k_{*}$. However, this may be
\emph{too} conservative, i.e, $k_{*} \gg N$, if a few edges are seen
to display a slow autocorrelation decay. If we are interested in
certain global metrics for  graph e.g., maximum eigenvalue etc, a few
correlated edges are unlikely to have any substantial effect. Thus,
one may be able to thin with a $k \sim N \ll k_{*}$. We will test this
empirically in Sec.~\ref{sec:tests}.

%% file: tests.tex
\section{Tests with real graphs}
\label{sec:tests}

In this section we first explore the impact of $\epsilon$ (as defined
in Sec.~\ref{sec:sesh}) on the ensemble of graphs generated by a
Markov chain, and choose a $\epsilon$ for further use. Thereafter we
compare the graphs generated by the Methods A and B
(Sec.~\ref{sec:sesh} and Sec.~\ref{sec:rl}) and gauge the impact of
choosing a thinning factor $k < k_{*}$. All tests are done with four
real networks - the neural network of \emph{C. Elegans}~\cite{celegans1}
(referred to as ``C. Elegans''), the power grid of the Western states
of US~\cite{celegans1} (called ``Power''), co-authorship graph of
network science researchers~\cite{adjnoun} (referred to as
``Netscience'') and a 75,000 vertex graph of the social network at
Epinions.com~\cite{03ra3a} (``Epinions''). Their details are in
Table~\ref{tab:graphs}. The first three were obtained
from~\cite{newman_graphs} while the fourth was downloaded
from~\cite{snap}. All the graphs were converted to undirected graphs by symmetrizing the edges.

We start the Markov chain using the real networks listed in
Table~\ref{tab:graphs}. When comparing ensembles of graphs, we will
use the (distributions of) global clustering coefficient, number of
triangles in the graphs, the graph diameter and the maximum eigenvalue
as metrics.

\begin{table}
  \begin{center}
    \caption{Characteristics of the graphs used in this paper. $(|V|,
      |E|)$ are the numbers of vertices and edges in the graph, $N$ are
      the number of Markov chain steps used for generating graphs in
      Sec.~\ref{sec:sesh} and $k$ is its equivalent obtained by the
      method in Sec.~\ref{sec:rl}. $K/k_*$ are the number of graph
      samples, obtained by thinning a long run, that were used to
      generate distributions in the figures.}
    \begin{tabular}{p{0.2\textwidth}p{0.32\textwidth}p{0.16\textwidth}p{0.16\textwidth}p{0.16\textwidth}} \hline
      \noalign{\smallskip}
      Graph name &  $(|V|, |E|)$   &   $N/|E|$   & $k_{*}/|E|$  & $K/k_{*}$ \\ \noalign{\smallskip}\hline   \noalign{\smallskip}
      C. Elegans & (297, 4296)     &     10      & 13       &   3582   \\ \noalign{\smallskip}
      Netscience & (1461, 5484)    &     10      & 49       &   737    \\ \noalign{\smallskip}
      Power      & (4941, 13188)   &     10      & 13       &   1214   \\ \noalign{\smallskip}
      Epinions   & (75879, 405740) &     30      & 720      &   various \\ \noalign{\smallskip}
      \hline \noalign{\smallskip}
    \end{tabular}
\label{tab:graphs}
  \end{center}
\end{table}

\begin{figure}[h]
  \centerline{\includegraphics[width=\textwidth, trim=2cm 7cm 2cm 7cm,clip=true]
   {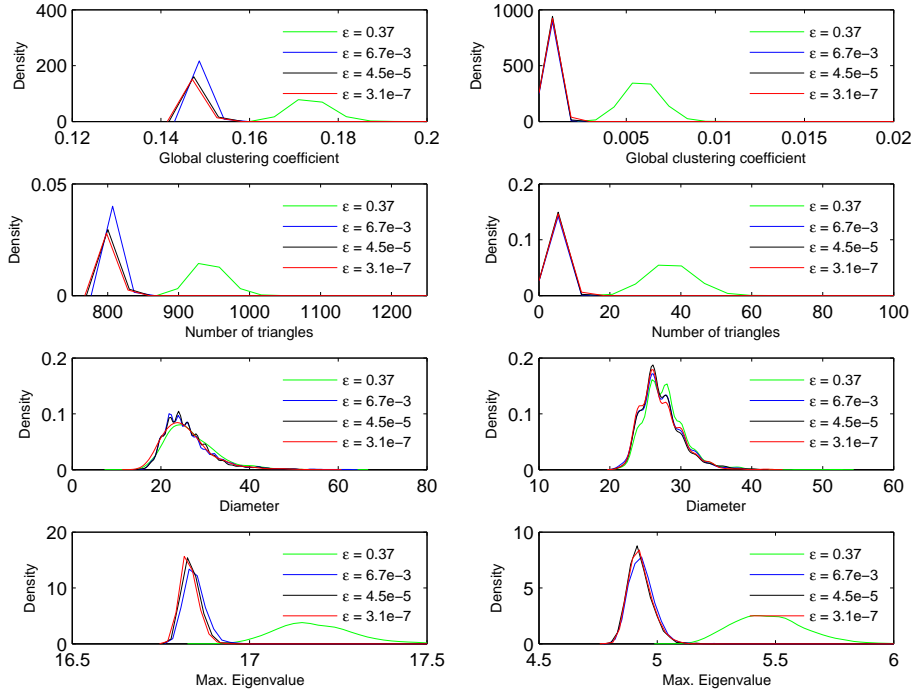}}
  \caption{ Plots of the distributions of the global clustering
  coefficient, the number of triangles in the graphs, the graph
  diameter and the max eigenvalue of the graph Laplacian for
  ``Netscience'' (left) and ``Power'' (right), evaluated after $1|E|,
  5|E|, 10|E|$ and $15|E|$ iterations of the Markov chain (green, blue,
  black and red lines respectively). The corresponding values of
  $\epsilon$ are in the legend. We see that the distributions converge
  at $\epsilon < 1.0^{-5}$.}
  \label{fig:epsilon}
\end{figure}

In Fig.~\ref{fig:epsilon} we investigate the impact of $\epsilon$ in
Method A (``many short runs''). We generate 1000 samples by running
the Markov chain for $1|E|, 5|E|, 10|E|$ and $15|E|$ Markov chain
iterations, corresponding to $\epsilon = 0.37, 6.7\times10^{-3},
4.5\times10^{-5}$ and $3.06\times10^{-7}$. In Fig.~\ref{fig:epsilon},
we plot the distributions for the first three graphs (in
Table~\ref{tab:graphs}) and find that for all three, $\epsilon < 5.0
\times 10^{-3}$ lead to distributions which are very close. We will
proceed with $\epsilon = 4.5\times10^{-5}$ i.e., when we use Method A,
we will mix the Markov chain $10|E|$ times before extracting a sample.

\begin{figure}[h]
  \centerline{\includegraphics[width=\textwidth, trim=2cm 7cm 2.cm 7cm,clip=true]
  {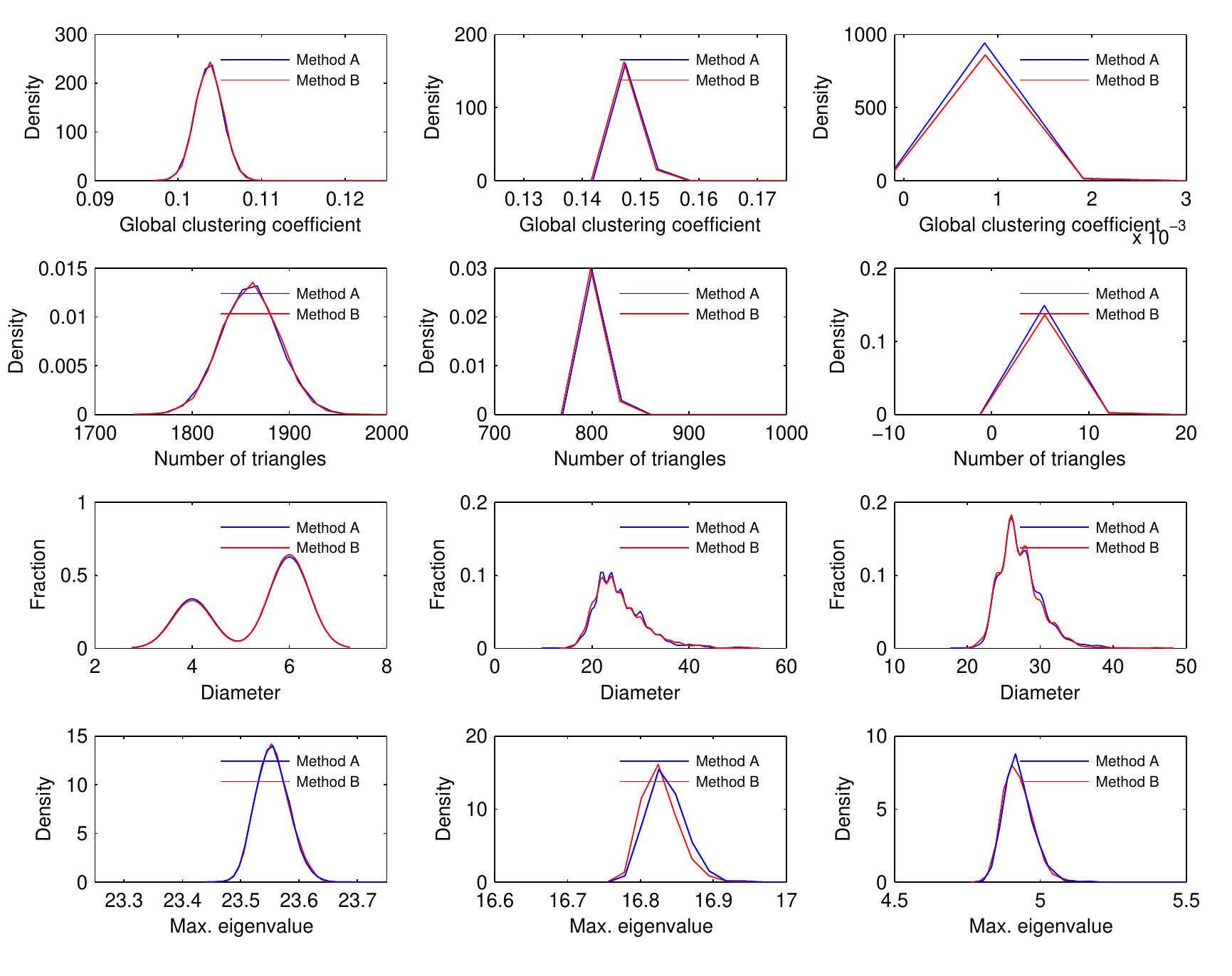}}
  \caption{Comparison of the distributions of the global clustering
   coefficient, the number of triangles in the graphs, the graph
   diameter and the max eigenvalue of the graph Laplacian for
   ``C. Elegans'' (left), ``Netscience'' (middle) and ``Power''
   (right), evaluated using Methods A (``many short runs'') and B
   (``one long run''). We see that the distributions are very
   similar. The kernel density estimation used to generate the
   distributions sometimes causes nonsensical artifacts e.g., a small,
   but negative clustering coefficient. For Method A, the Markov chain
   was run for $10|E|$ iterations. Thinning factors for Method B are in
   Table~\ref{tab:graphs}. }
  \label{fig:ab}
\end{figure}

In Table~\ref{tab:graphs} we see that Method B (``one long run'')
method often prescribes a thinning factor that is larger than the one
obtained using Method A (``multiple short runs''). This large number
is often due to the lack of autocorrelation decay in a few edges. We
investigate whether such a lack has a significant impact on the
graphical metrics that we have chosen. In Fig.~\ref{fig:ab} we plot
distributions of the same metrics for the three graphs. The thinning
factors are in Table~\ref{tab:graphs}. We see that the distributions
are close, i.e., the existence of a few edges whose time-series are
still correlated do not impact the metrics of choice. We have repeated
these tests with other metrics and the same result holds true.

\begin{figure}
  \centerline{
    \includegraphics[width=0.5\textwidth, trim=1cm 6cm 2.cm 7cm,clip=true]{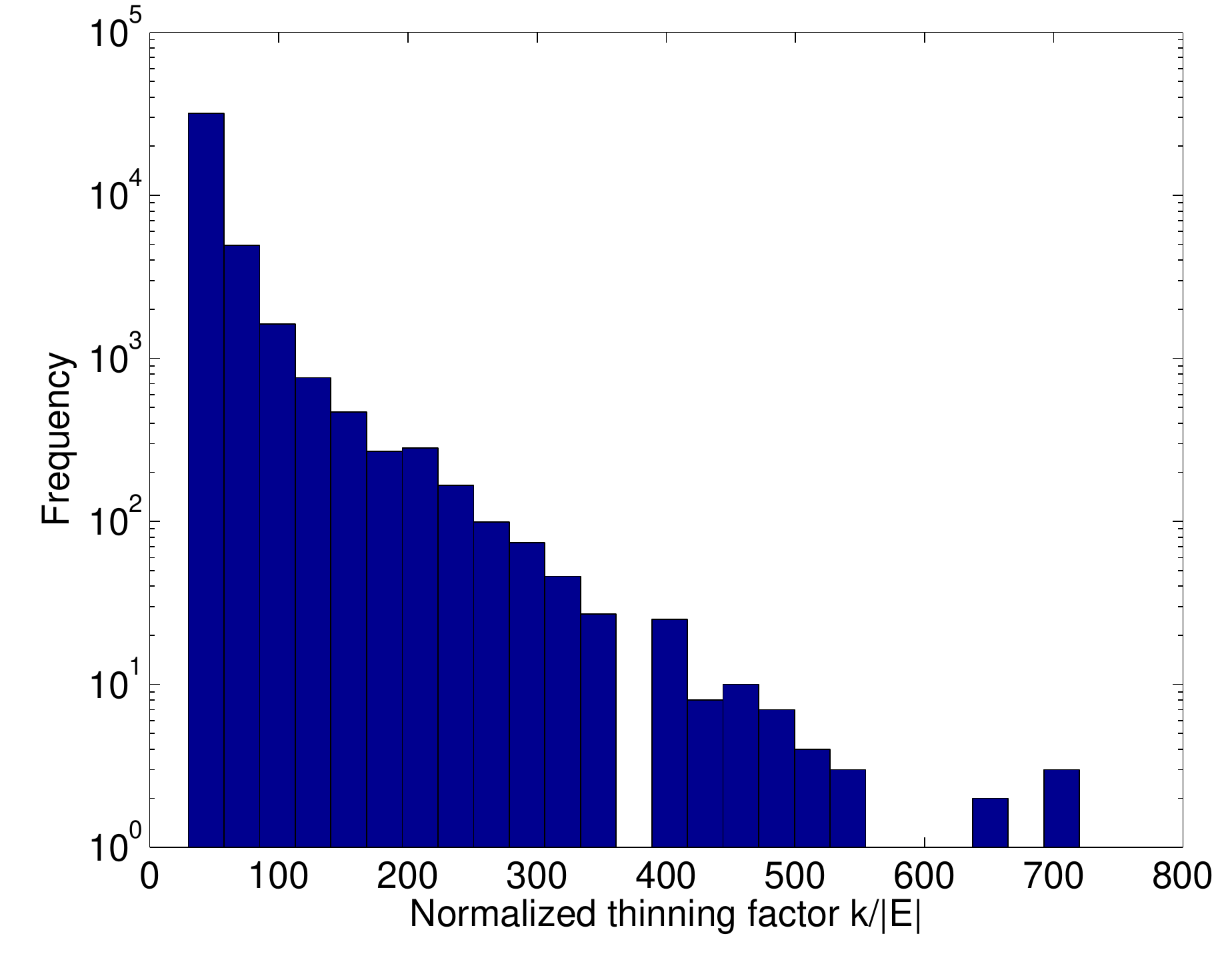}
    \includegraphics[width=0.5\textwidth, trim=1cm 6cm 2.cm  7cm,clip=true]{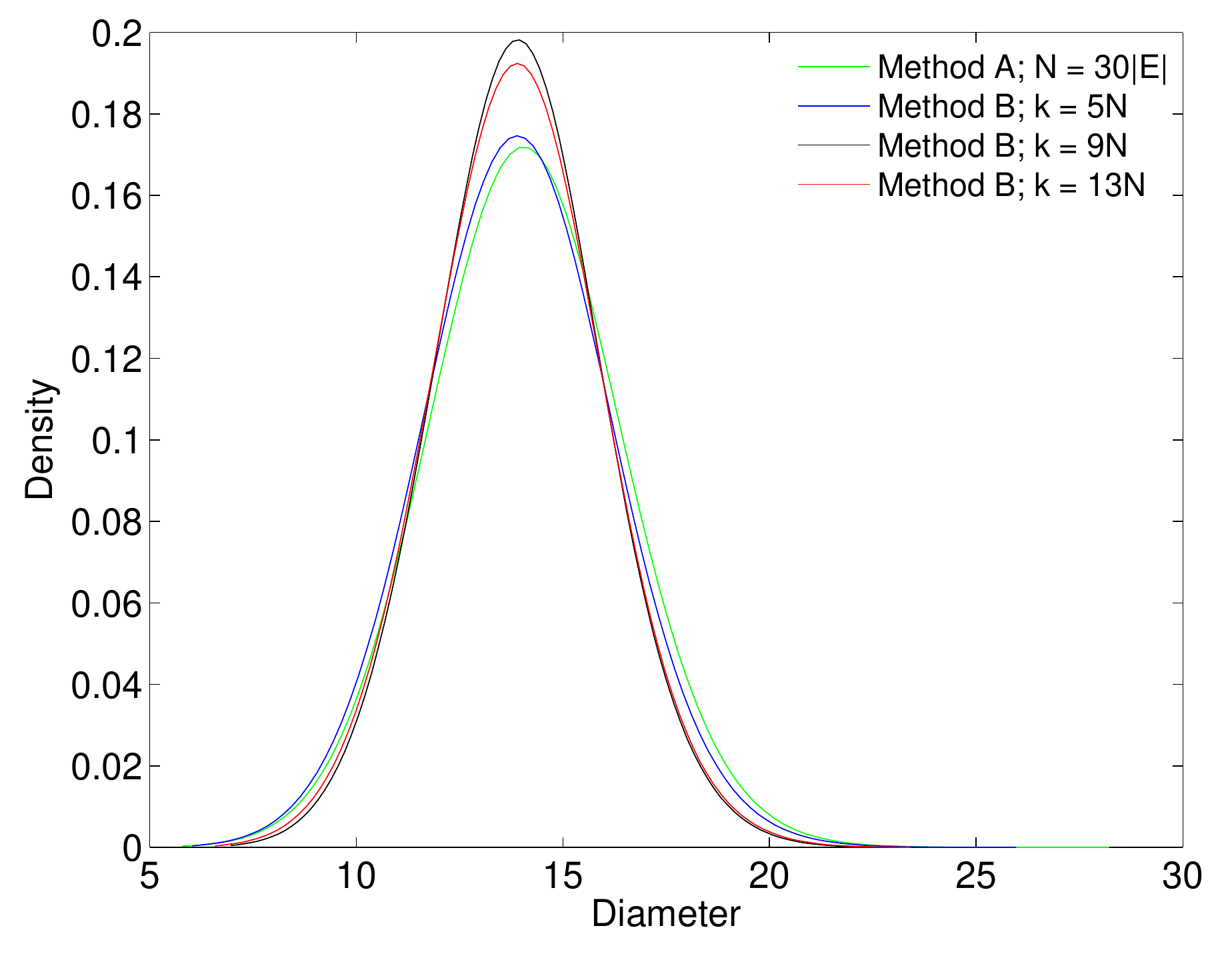}
  }
  \caption{Left: The normalized thinning factor $k/|E|$ for the
  Epinions graph, as calculated for the 40,574 sampled edges. We see
  that the most thinning factors are lie in ($10|E|, 100|E|$).  Right:
  Plot of the graph diameter and distribution generated using Method A
  (with $N = 30|E|$) and Method B (with $k$ equal to various multiples
  of $N$). We see that the distributions are very similar.}
  \label{fig:k}
\end{figure}

We now address a large graph (Epinions). Since potentially $|V|^2$
distinct edges might be realized during a Markov chain, it is
infeasible to calculate a thinning factor for all the
edges. Consequently, we perform the thinning analysis for only
$0.1|E|$ (40,574) edges, chosen randomly from all the distinct edges
that are realized by the Markov chain. In Fig.~\ref{fig:k} we plot the
distribution of $k$ obtained from the 40,574 sampled edges. We see
that most of the $k$ lie between $10|E|$ and $100|E|$; edges with
thinning factors outside that range are about two orders of magnitude
less abundant. It is quite conceivable that there are edges (which
were not captured by the sample) that would prescribe an even higher
thinning factor. In order to check whether these edges have a
significant impact on the distribution of graph metrics, we check
their convergence as a function of the thinning factor.

We generate separate ensembles of graphs.  The reference ensemble is
generated using Method A, with $N = 30|E|$. As seen in
Table~\ref{tab:graphs}, certain edges will still be correlated ($N =
720|E|$ would make them independent). We then use Method B to generate
graph ensembles with thinning factors $k/|E| < 720$ which are also
multiples of $N$. In Fig.~\ref{fig:k} (right), the diameter
distribution obtained with Method A is compared to that obtained with
Method B. While the distributions are very similar, they do display
some small differences. This is surprising since $N = 30|E|$ indicates
a minuscule $\epsilon$. In addition, distributions obtained with $k =
5N, 9N,$ and $13N$ show some differences between themselves, indicating
that the edges that have not become independent have a small, but
measurable impact on the graph diameter. Further, the distributions
using smaller values of $k$ are marginally wider (have a larger
variance), indicating that they were constructed using samples which
were not completely independent. However, the differences are minute,
and for practical purposes the graph ensemble generated using Method A
with $N = 30|E|$ is identical to the one generated using Method B, per
our chosen metrics. Consequently, despite its approximations, the
results in Sec.~\ref{sec:sesh} furnish a workable estimate of $N$, if
one uses $\epsilon < 10^{-5}$. Further, $k_{*}$ is generally too
conservative if our aim is to obtain ``converged'' distributions of
certain graph metrics. This arises from a few edges that
de-correlate slowly, but have little effect on global graphical
metrics due to their rarity.

%% file: concl.tex
\section{Conclusions}
\label{sec:concl}

We have developed a method that allows one to generate a set of
independent realizations of graphs with a prescribed joint degree
distribution. The graphs are generated using an MCMC approach,
employing the algorithm described in~\cite{StPi12} as the ``rewiring''
mechanism. The graphs so generated are tightly correlated; our two
methods address the question of how one can decorrelate the chain.

The first method, variously called Method A or ``multiple short
chains'', involves running the Markov chain for $N$ steps before
extracting a graph realization; the Markov chain is run repeatedly to
generate samples. We developed a model (and a closed-form expression)
to estimate $N$ that allows the Markov chain to converge to its
stationary distribution before a graph realization is extracted from
it. This model assumes that edges are independent. In reality, their
behavior is correlated, which leads us to incur small errors.

The second method, variously called Method B or ``one long chain'',
is a data driven method. It uses the time-series of the
occurence/non-occurence of edges in an MCMC run. It does not assume a
constant joint degree distribution. It progressively thins the
time-series (by retaining every $k^{th}$ element) and fits a
first-order Markov and an independent sampling model to the data. The
thinning process stops when the independent model has a higher
likelihood (strictly, a lower BIC score) than the Markov process. Since this
method is data-driven and does not require any user-defined
tolerances, we use it to validate Method A. The method is not new, but
does not seem to have been used in the generation of independent
graphs.

Comparing the two methods, we find that for practical purposes, the
ensembles generated using Method A are statistically similar to those
obtained with Method B, as gauged by a set of graph metrics. Even at
tight tolerance values, a small number of edges in the graphs
generated by Method A remain correlated, and the metrics'
distributions are slightly wider. This problem is very small (nearly
unmeasureable) in small graphs, but becomes measureable, but still
small, for large graphs.

While this work enables the generation of independent graphs,
including large ones, it poses a number of questions for further
investigation. For example, being able to estimate or bound the
difference in the distributions generated by Methods A and B would be
helpful. Further, an intelligent way of identifying
hard-to-decorrelate edges would reduce the computational burden of
checking for the stopping criterion using Method B; currently, we
simply use a random set of edges. Finally, it would be interesting if
Method A could be extended to the generation of independent graphs
when some graph property, other than the joint degree
distribution, is held constant. This is currently being studied.